**Title:** Magneto-structural correlations in doubly hydroxo-bridged Cu(II)-dimers

**Author:** Stefan Lebernegg

**Address:** Department of Materials Engineering and Physics, University of Salzburg, Hellbrunnerstrasse 34, 5020 Salzburg, Austria

**Running title:** Magnetism of di-µ-hydroxo-bridged Cu-dimers

**Corresponding authors address:**

Phone:  ++43-662-8044-5463

Fax: ++43-662-8044-622

Mail: stefan.lebernegg@sbg.ac.at

**Abstract.** The magneto-structural correlations of superexchange-coupled doubly hydroxo-bridged Cu(II)-dimers have been investigated. To this end, an analytical approach has been applied to $[Cu_2(OH)_2F_4]^{2-}$ model complexes. This approach supplies an analytical scheme, based on orbital interactions, for calculating the transfer integral, $H_{AB}$, which is shown to play the key role in the magnetic coupling constant $J$ for understanding magneto-structural correlations. The single contributions to the transfer integral are calculated and described explicitly. Therefore, this approach supplies a detailed insight into the magnetic behavior and the interaction mechanisms of the hydroxo-bridged Cu-dimers. All analytical results are compared with experimental and numerical data.

## 1. Introduction

The doubly hydroxo-bridged Cu(II) dimers are one of the most extensively experimentally studied molecular magnets [1,2]. Since Cu(II) is a $d^9$ ion there is merely one unpaired electron (active electron) per metal center. Therefore, these complexes represent a relatively simple system for describing magneto-structural correlations in superexchange-coupled compounds. Beside the dependence of the magnetic coupling on the bridging angle, its correlations to the out-of-plane angle of hydrogen, the copper-oxygen bonding-distance and the bending angle of the molecules have been investigated both experimentally [3,4,5] and theoretically [1,6,7]. Additionally, empirical relations between the magnetic coupling constant $J$ and geometrical parameters were derived from experimental data[3] without, however, giving any insight into the coupling mechanisms. Numerical calculations of $J$ with density functional theory (DFT) could reproduce the experimental results and, based on calculations on model systems, they supplied important information about the relations between geometrical parameters and magnetic coupling [1,6]. An analysis and an interpretation of the magnetic behavior were usually based on the expression for the magnetic coupling constant derived by Hay et al. [7]

$$J = K_{AB} - \frac{2(H_{AB})^2}{U} \qquad (1)$$

where the two terms on the right side give the ferromagnetic and antiferromagnetic contributions, respectively. $K_{AB}$ is a two-electron exchange integral, $U = U_{AA} - U_{AB}$ (the effective Hubbard $U$) is the difference between one- and two-center two-electron Coulomb-integrals and $H_{AB}$ is the transfer integral that can be obtained from the numerical calculations (e.g. refs.[1,7]). $H_{AB}$ is proportional to the probability of the electron to hop from one centre of the dimer to the other and is nonzero only for antiparallel spin alignment of the two active electrons of the dimers. The two-electron integrals are commonly assumed to be insensitive to structural changes. Thus, $J$ should essentially depend on the square of $H_{AB}$ [8] where a large HAB favors antiferromagnetic coupling. However, the mechanisms and interactions determining HAB, that can hardly be extracted from the output of numerical methods, remain hidden. A detailed and quantitative analysis of the single contributions to the transfer integral that would supply a better understanding of the magnetic behavior is still missing. Therefore, the aim of this work is to investigate the magneto-structural correlations in the doubly hydroxo-bridged Cu(II)-dimers in more detail based on an orbital interaction scheme. To this end, an approach [9] is applied that enables the analytical calculation of the transfer integral $H_{AB}$ which was explicitly shown to be the crucial quantity for the magneto-structural correlations. $H_{AB}$ is represented as a function of the energies of metal and ligand orbitals and the overlap integrals. In the present work this analytical formalism is applied on $[Cu_2(OH)_2F_4]^{2-}$ model-complexes where the bulky terminal ligands of the real systems (e.g. ref.[2,3]) are substituted by fluorine enabling an easier analysis. Since the structural variations discussed here in principle only affect the interactions between Cu and OH, the magneto-structural correlations should be qualitatively independent of the choice of the terminal ligands, as long as the distorted square-planar coordination of copper is preserved. The magnetic behavior will be investigated for variations of the following structural parameters: bridging angle $\theta$, out-of-plane angle of hydrogen $\tau$, bending angle $\gamma$ (bending about the axis parallel to the bridging (OH)-molecules), copper-oxygen bonding distance. All analytical calculations are supplemented by fully numerical calculations with the self-consistent charge (SCC)-X$\alpha$ program [10,11].

2. **Computational details**

For the planar $[Cu_2(OH)_2F_4]^{2-}$ model complex, the following structural parameters are used, if not stated otherwise: Cu-O = 1.95Å, Cu-F = 1.94Å, O-H = 0.96Å, F-Cu-F = 93°. The *xz*-plane is chosen as the molecular plane, where *z* is the internuclear axis of the two monomers (figure 1).

**Fig. 1.**

The transfer integral $H_{AB}$ may be defined as[7]

$$H_{AB} = \frac{1}{2}(\varepsilon_+ - \varepsilon_-) \qquad (2)$$

where $\varepsilon_\pm$ are the orbital energies of the magnetic molecular orbitals of the dimer that contain the active electrons. These molecular orbitals (MOs) belong to a symmetrical (-) and an unsymmetrical (+) linear-combination of atomic orbitals (AOs). As basis set for the analytical calculations Cu(3$d$), F(2$s$,2$p$), O(2$s$,2$p$) and H(1$s$) AOs are chosen which are described by the product of a single Slater-type orbital (STO) and a real spherical harmonic. The orbital exponent can be taken as constant for different geometries [9]. The 4$s$ and 4$p$ orbitals of copper are assumed to be negligible.

For the analytical calculation of the magnetic MOs and their energies $\varepsilon_\pm$ it is convenient to construct two sets of group-orbitals from the AOs of the basis set according to the symmetries of the two magnetic MOs. The analytical calculation of $H_{AB}$ is performed in general in two steps [9,12] aiming to transform the full multi-center MO-Hamiltonian into an effective single-center problem. First, the Kohn-Sham equation is solved in linear combination of atomic orbital approximation with respect to Cu(3$d$) group orbitals that are Schmidt-orthogonalized to the ligand group orbitals. The latter are assumed to be orthogonal or have been orthogonalized among each other. In the second step, the Hamiltonian-matrix, calculated with the orthogonalized orbitals, is diagonalized analytically. This matrix is of block-diagonal form due to the use of the symmetry-adapted basis. This procedure was denoted as dimer approach [9]. In combination with a bridging-ligand-only method [9], where matrix elements of the terminal ligands are removed from the block-diagonalized Hamiltonian matrix, it leads to sufficiently low dimensional blocks that can be diagonalized analytically. Those eigenfunctions having predominantly Cu(3$d$) character correspond to the magnetic orbitals and their eigenvalues correspond to $\varepsilon_\pm$. These energies cannot be approximated via a perturbation calculation since some nondiagonal Hamiltonian matrix-elements are too large. The neglect of the terminal ligands in the diagonalization step is well justified in these complexes since the interactions between fluorine and copper are quite weak and cancel each other to a very high degree in the energy-difference of eq.(2).

For an analytical solution the orbital energies of the AOs and the overlap integrals and the nondiagonal Hamiltonian matrix between the AOs are required. The first have turned out to be more or less constant with respect to geometrical changes of the dimers and are taken to be the orbital energies of the atoms in the molecule. The overlap integrals for STOs can be calculated analytically [13]. For the nondiagonal Hamiltonian matrix elements the following approximation is used [12]:

$$H_{mn}^{XY} = \frac{1}{2}\left(H_{mm}^{XX} + H_{nn}^{YY} + 2\bar{v}_{mn}^{XY}\right) \cdot S_{mn}^{XY} \quad (3)$$

where $m,n$ are any AOs on the centers $X,Y$. $H_{mm}^{XX}$ and $H_{nn}^{YY}$ are the diagonal elements of the Hamiltonian matrix and are approximated as orbital energies of atoms in the molecule. $\bar{v}_{mn}^{XY}$ is an angular-independent averaged potential of 2- and 3-centre Coulomb and exchange integrals.

The H(1$s$)-orbitals contribute solely to the symmetrical magnetic MO and interact strongly with the O(2$s$) and O(2$p_x$) orbitals. In order to include these interactions it is advantageous to calculate first the eigenfunctions of the OH$^-$ molecule with O(2$s$), O(2$p_x$) and H(1$s$) as basis. If the molecule is not planar it is necessary to include also the O(2$p_y$)-orbital (see below). These eigenfunctions together with the group orbitals from the terminal ligands are the proper ligand-orbitals to which the Cu(3$d$) group orbitals have to be orthogonalized for the symmetric MO. If not explicitly stated otherwise, in the following discussions orbitals are always understood as group-orbitals for Cu(3$d$), O(2$p_z$) and the terminal ligands and as eigenfunctions of the OH$^-$ molecule for the bridging orbitals contributing to the symmetric magnetic MO. In order to simplify the calculations it is assumed that all interactions between ligands but those between oxygen and hydrogen can be neglected. Only for the variation of the Cu-O bonding distances it will turn out that orthogonalization of the $p_z$-orbitals of bridging and terminal ligands may significantly improve the results. Consequently, the only nondiagonal elements in the Hamiltonian-matrix are those between the orthogonalized Cu(3$d$) and the ligand orbitals.

The analytical results are compared with fully numerical calculations with the SCC-X$\alpha$ method whereas the numerical transfer integrals are taken from spin-restricted calculations as proposed by Seo [14], i.e. $H_{AB}$ corresponds to the half of the HOMO-LUMO gap. The magnetic coupling constant $J$ is calculated numerically in local spin-density approximation in combination with the broken-symmetry formalism [15,16].

$$J(bs) = -\left[E\left(S^{max}\right) - E\left(S^{BS}\right)\right]/\left(S^{max}\right)^2 \quad (4)$$

where $E(S^{max})$ and $E(S^{BS})$ correspond to the energies of the ferromagnetic and broken symmetry states, respectively, and $S$ is the total spin. For the analytical calculations of $J$ a parameterized form of eq.(1) is used[9]

$$J = C - \frac{2(H_{AB})^2}{f} \qquad (5)$$

where $C$ and $f$ are constant parameters corresponding to the direct exchange and the Hubbard $U$, respectively. Inserting the analytical $H_{AB}$ into eq.(5) and choosing two reasonable values for $C$ and $f$ should supply a good description of $J$ as a function of geometrical parameters. For comparison also numerical transfer integrals will be inserted.

3. Results

**Planar complex with varying bridging angle $\theta$**

In a first step the eigenvectors and eigenvalues of the OH⁻ molecule are calculated. Due to symmetry only the O($2s,2p_x$) and H($1s$) orbitals contribute. Two of the three eigenfunctions are separated by more than 20eV from the Cu($3d$)-orbital, which is of pure $d_{xz}$ character, so that only one orbital should contribute significantly to the symmetric magnetic MO. This orbital has the character of about 50% O($2p_x$) and 25% of each O($2s$) and H($1s$) The transfer-integrals are calculated analytically for bridging angles between 80° and 110°. The results are shown in figure 2 together with those from SCC-Xα calculations. For comparison contributions from ligand-ligand interactions (LLI) are added to the analytical values. These interactions slightly reduce the slope of the transfer integral becoming very similar to the fully numerical results. The most important contributions arise for small bridging angles from the interactions between the $p_z$-orbitals of bridging and terminal ligands while for large angles especially interactions between bridging AOs of oxygen come into play. However, with respect to the simplifications of the analytical description this error is acceptable.

**Fig. 2.**

The great advantage of the analytical calculation compared with, e.g. DFT-calculations, is the separability of the single contributions to $H_{AB}$. This was already demonstrated for Cu-F complexes [9]. The expressions for the single contributions to $H_{AB}$ applied there have been somewhat adapted for the dimer approach of complexes with stronger orbital interactions as this is the case for the hydroxo-bridged dimers. In order the keep the expressions handy and simple, some small terms entering into to calculation of $\varepsilon_{\pm}$ described above are neglected so that the sum of all contributions may deviate slightly from the total analytical value of $H_{AB}$. The contribution from the direct $d$-$d$ interaction is

$$c(dd) = 2\left(H_{dd}^{AB} - 2 \cdot S_{dd}^{AB} \cdot H_{dd}\right) \qquad (6)$$

where $H_{dd}^{AB}$ and $S_{dd}^{AB}$ are Hamiltonian and overlap matrix elements between atomic Cu(3$d$)-orbitals on different centers of the dimer. $H_{dd}$ is the energy of the atomic Cu(3$d$)-orbital. The contributions from the interactions between the Cu(3$d$) and bridging ligand orbitals are given as

$$c(kd^{\pm}) = \pm \frac{1}{2}\left(S_{kd}^{\pm,2} \cdot H_{dd} - 2H_{kd}^{\pm}S_{kd}^{\pm} + S_{kd}^{\pm,2}H_{kk} + dia^{\pm}\right) \qquad (7)$$

where $\pm$ denotes contributions to the symmetrical (-) and unsymmetrical (+) magnetic MOs, respectively. $H_{kd}$, $S_{kd}$ are nondiagonal elements between Cu(3$d$) and bridging ligand orbitals of the Hamiltonian and overlap matrix and $H_{kk}$ is the energy of a bridging ligand orbital. Small contributions from diagonalization (e.g. from s- and H-eigenfunctions) may be included as perturbations

$$dia^{\pm} = \frac{\tilde{H}_{kd}^{\pm,2}}{\left(H_{dd} - H_{kk}\right)}, \quad \left|\tilde{H}_{kd}\right| \quad \left|H_{dd} - H_{kk}\right| \qquad (8)$$

whereas large ones are taken from direct diagonalization

$$dia^{\pm} = \frac{H_{kk} - \tilde{H}_{dd}^{\pm}}{2} + \frac{1}{2}\left(\left(\tilde{H}_{dd}^{\pm} - H_{kk}\right)^2 + 4\tilde{H}_{kd}^{\pm}\right)^{1/2} \qquad (9)$$

$\tilde{H}_{dd}$ is the energy of the ligand-orthogonalized Cu(3$d$)-orbital and $\tilde{H}_{kd}$ its matrix element with a bridging ligand orbital. The single contributions from the bridging ligands are listed in table 1 whereas for $c(xd)$ and $c(zd)$ eq.(9) is used.

**Table 1.**

It has to be kept in mind that the bridging ligand orbitals contributing to the symmetrical magnetic MO are eigenfunctions of the OH⁻ molecule so that $s$, $x$, $H$ denote their dominant character where $s$ = O(2$s$), $x$ = O(2$p_x$) and $H$ = H(1$s$). The first column in table 1 gives the contribution from the direct $d$-$d$ interaction that decreases with increasing $\theta$ due to the increasing Cu-Cu distance. The $s$-$d$ interaction remains more or less constant since the O(2$p_x$) contribution to the $s$-eigenfunction is very small. Consequently, there is only a marginal angular dependence of the interaction with the Cu(3$d$) orbital. The distance between hydrogen and copper is too large so that this interaction is negligible. Therefore, the magneto-structural correlations are dominated by the $z$-$d$ and $x$-$d$ interactions which cancel each other at about 85°, in qualitative agreement with the empirical Goodenough-Kanamori rules [17,18]. Due to an admixture of 25% O(2$s$) to the $x$-eigenfunction this compensation does not occur at exactly 90°.

Experimental and numerical data for the magnetic coupling constant of doubly hydroxo-bridged Cu-dimers as a function of the bridging angle exhibit a transition from antiferromagnetic to ferromagnetic coupling at about 90° [1] and at about 97° [3,7], respectively. The broken symmetry calculations with SCC-Xα yield such a transition at about 92° (figure 3). A reduction of the bridging angle leads again to antiferromagnetic coupling. When inserting the analytical and for comparison also the numerical transfer integrals (see figure 2) into eq.(5) the magnetic behavior is well reproduced over the whole range of $\theta$.

**Fig. 3.**

These results confirm that the magneto-structural correlations for varying bridging angles are fully determined by $H_{AB}$ and its dependence on the $p$-$d$ interactions corresponding to what one would expect from the Goodenough-Kanamori rules.

**Complex with rotated H-atoms**

For these calculations the $[Cu_2(OH)_2F_4]^{2-}$ complex with a bridging angle of 95° is taken where no significant contributions from the LLI are expected (see figure 2). Starting with the planar complex the hydrogen atoms are rotated into opposite directions out of the molecular plane by an angle $\tau$. Experimental and theoretical data [1,6,19] show that successive increasing of this angle leads to a transition from antiferromagnetic to ferromagnetic coupling.

The procedure for the analytical calculation is in principle the same as in the previous chapter with the only difference that additionally the $O(2p_y)$-orbital contributes to the symmetrical magnetic MO due to the reduced symmetry, i.e. there are now four $OH^-$ eigenfunctions. Two of them strongly interact with $Cu(3d)$, i.e. the block of the Hamiltonian matrix corresponding to the symmetric magnetic MO is three-dimensional (the two small contributions may be considered as perturbations or can be neglected). The eigenvalues may be approximated analytically with sufficient accuracy by separating the three-dimensional into two two-dimensional problems. The analytical transfer integrals for different angles $\tau$ are displayed in figure 4 together with the numerical values.

**Fig. 4.**

There are only very small deviations between the numerical and the analytical transfer integrals both decreasing with increasing $\tau$. The single contributions to $H_{AB}$ calculated with eqs.(6)-(9) are given in table 2. The $x$- and $y$-eigenfunctions are supplemented by a prime indicating that their characters change with increasing $\tau$.

**Table 2.**

The direct $d$-$d$ and the $z$-$d$ interactions are constant since they are not affected by the position of hydrogen. The interaction between $Cu(3d)$, which is again of pure $d_{xz}$ character, and the $s$-eigenfunction is more or less constant, too, since this eigenfunction has predominantly $O(2s)$ character and the interaction of $O(2s)$ with $H(1s)$ is independent of the hydrogen angle. The weak increase of its absolute value arises from slightly varying contributions from the O(2$p$)-orbitals. The interaction between the *H*-eigenfunction and Cu(3$d$) is again negligible and decreases due to the decreasing overlap between the Cu(3$d$) and H(1$s$) orbitals for increasing $\tau$. Thus, the magneto-structural correlations are fully described by the *y'-d* and *x'-d* interactions. The *y'*-eigenfunction of the OH$^-$ molecule has negligible contributions from the H(1$s$) and O(2$s$) orbitals. In particular, for the planar complex it is a pure O(2$p_y$)-orbital not interacting with Cu(3$d$). When rotating the hydrogen atom, the O(2$p_x$)-character of the *y'*-eigenfunction increases and consequently its interaction with Cu(3$d$). Finally, for angles larger than 45° the O(2$p_x$)-character dominates. The opposite happens to the *x'*-eigenfunction, where the O(2$p_y$)-character increases with increasing $\tau$ accompanied by a decreasing interaction with Cu(3$d$). While the *y'*-function has no contributions from O(2$s$) and H(1$s$)-orbitals, their contributions to *x'* make almost 50%, hence, the absolute increase of the O(2$p_x$)-character in the *y'*-function is larger than its decrease in the *x'*-function. In summary, the variations in strength of the *y'-d* and *x'-d* interactions do not compensate but give a net negative contribution to $H_{AB}$ responsible for its decay with increasing $\tau$.

The curves obtained from inserting the analytical and numerical $H_{AB}$ into eq.(5) are shown in figure 5 together with the results from the broken symmetry calculations.

**Fig. 5.**

The decreasing strength of the antiferromagnetic coupling and finally the transition to the ferromagnetic state is consistent with both other numerical calculations and experimental results. The agreement between the broken symmetry results and the coupling constants calculated with eq.(5) is satisfying again and demonstrates that the analytical approach is capable of describing the magnetic behaviour and the orbital-interaction analysis given above supplies a correct description of the magnetic coupling mechanism. Moreover, this magnetic behavior agrees with the results from Ruiz et al.[1] who observed a decrease of the antiferromagnetic coupling with increasing $\tau$. For a bridging angle $\theta = 95°$ a transition to ferromagnetic coupling was predicted for an out-of-plane angle of about 40°. In our calculations (see figure 5) this transition occurs at about 30° which may be traced back to the different choice of the terminal ligands.

**Bent complex**

Experimental and theoretical data reveal a decreasing strength of the antiferromagnetic coupling with increasing bending angle $\gamma$ [1,4,5], finally leading to a transition to ferromagnetic coupling. For the investigation of the bent dimers again the model-complex with a bridging angle of 95° is used. The monomers of this complex will then be rotated into opposite directions by an angle $\gamma$ about the axis connecting the OH-bridges (figure 6). The dihedral angle of the dimer is defined as 180°-2$\gamma$. The calculation procedure is the same as in the previous case. However, the monomers are not in the *xz*-plane, so that, the Cu(3*d*)-orbital is not anymore of pure $d_{xz}$-character but is a linear combination of $d_{xz}$ and $d_{yz}$. As discussed in ref.[9] for the singly bridged $[Cu_2F_7]^{3-}$ the coefficients of the linear combination should be estimated first, i.e. before the orthogonalization and diagonalization steps, so that the analytical solvability of these steps is preserved. A simple approximation of this linear combination would be to rotate the $d_{xz}$-orbitals of both monomers into the respective monomer-planes giving the Cu(3*d*)-orbital as

$$d_\gamma = \cos(\gamma) \cdot d_{xz} \pm \sin(\gamma) \cdot d_{xy} \tag{10}$$

The sign of the $d_{xy}$-contributions is (-) for the monomer A and (+) for monomer B (figure 6).

**Fig. 6.**

This approximation supplies the best results when the monomers interact only via the bridging ligands and the linear combinations are similar for both magnetic MOs. Both conditions are sufficiently fulfilled in the discussed model-system. The numerical and analytical results for the transfer integral are shown in figure 7.

**Fig. 7.**

The transfer integrals strongly decrease with increasing bending angle. The single contributions to the analytical $H_{AB}$ are listed in table 3.

**Table 3.**

Due to the bending the Cu-Cu distance is reduced with corresponding increase of the *d-d* interaction. The *s-d*, *H-d* and *x-d* contributions are approximately constant for symmetry reasons, though some slight deviations arise from the approximation eq.(10). Due to the rotation out of the *xz*-plane the strength of the interaction between Cu(3*d*) and the *y*-eigenfunction, which is of pure O(2$p_y$) character strongly increases on expense of the *z-d* interaction. Since $c(zd)$ is positive and $c(yd)$ is negative this results in a negative contribution to $H_{AB}$ that is reduced by the direct *d-d* interaction. Inserting the transfer integrals into eq. (5) yields the coupling constants shown in figure 8.

**Fig. 8.**

All curves behave similarly up to about $\gamma = 15°$ and exhibit the expected decrease of the antiferromagnetic coupling strength. For larger angles the decay into the antiferromagnetic region of the analytical coupling constant is too strong due to the too large slope of $H_{AB}$ (figure 7). This may be traced back to the approximation eq.(10) since it assumes that the linear combinations are the same for both magnetic MOs. Indeed, the linear combinations become successively different for increasing $\gamma$ basically due to the different types of bridging orbitals contributing to these MOs. Nevertheless, the analytical calculation supply a correct description of the transition from antiferromagnetic to ferromagnetic coupling and predict at least qualitatively the return to an antiferromagnetic behavior for large bending angles.

In the work of Ruiz et al.[1] the first transition appears at about 20° for a bridging angle of 95°. figure 8 shows this transition at about 13°. The difference may again be assigned to the different terminal ligands. Charlot et al.[4,5] explain this transition with a stabilization of the antisymmetric MO with increasing $\gamma$. The symmetric MO should be unaffected by the structural variations and since it is lower in energy the energy difference to the antisymmetric MO and thus $H_{AB}$ decreases leading to ferromagnetic coupling. Our results yield a similar stabilization of the antisymmetric MO due to the decreasing $z$-$d$ interaction (table 3). However, due to the increasing strength of the $y$-$d$ interaction the symmetric orbital does not remain constant in energy but is destabilized. A constant energy is obtained only if neglecting the contribution from the $y$-eigenfunction.

If complexes with bridging angles of about 90° are bent one would start with a ferromagnetic complex (see figure 3) and would observe a transition to antiferromagnetic coupling. For even smaller bridging angles only an increasing strength of the antiferromagnetic coupling would be obtained.

**Variation of the Cu-O bonding-distance**

Numerical calculations and experimental results [1,3] reveal a transition from antiferromagnetic to ferromagnetic coupling when reducing the Cu-O distance which is attributed to an increase of the direct ferromagnetic exchange [1]. If this were indeed the case, the parameterized formula for $J$, eq.(5), cannot be applied.

For the analytical calculations the planar complex $[Cu_2(OH)_2F_4]^{2-}$ is chosen with a bridging angle fixed at 95°. The Cu-O bonding distance is varied between 1.70 and 2.20 Å. The calculation procedure for this planar complex is analogous to that for varying bridging angles with the only difference that terminal and bridging $p_z$-orbitals are orthogonalized to each other. This has turned out to be important for small bonding distances. Fully numerical and analytical transfer integrals are compared in figure 9.

**Fig. 9.**

Numerical and analytical transfer integrals show a similar decay for decreasing bonding distances indicating a decreasing antiferromagnetic coupling. If an improvement of the analytical results is desired for very small bonding distances, especially the nondiagonal elements between terminal and bridging ligands have to be included properly even in the diagonalization step. The single contributions to the analytical $H_{AB}$, using eqs.(6)-(9), are given in table 4.

**Table 4.**

As expected the absolute values of all interactions decrease with increasing bonding distance. However, the sum of the *z-d* and *x-d* interactions is approximately constant. Therefore, the magneto-structural correlations are not determined by the *p-d* interactions. The increase of $H_{AB}$ (figure 9) with increasing Cu-O distance is caused by the strong decrease of the absolute value of the *s-d* interaction that is somewhat compensated by the direct *d-d* interaction. The importance of the *s-d* interaction when varying the bonding distance was already demonstrated for a planar doubly-bridged $[Cu_2F_6]^{2-}$ complex[9] where eq.(5) has turned out to be applicable. Applying eq.(5) on the hydroxo-bridged complex yields the coupling constants displayed in figure 10.

**Fig. 10.**

Aside from some deviations at large bonding distances, the coupling constants from broken symmetry calculations can perfectly be reproduced with the numerical $H_{AB}$ confirming that eq.(5) is applicable. Therefore, the driving force for the transition from antiferromagnetic to ferromagnetic coupling is the decrease of the transfer integral (figure 9) while the ferromagnetic term and also the effective Hubbard U may still be assumed as constant. The analytical calculations are again in very satisfying agreement with the numerical results and offer a simple description of the magneto-structural correlations (table 4). Only for very small bonding distances some deviations occur due to the increasing strength of the ligand-ligand interactions (see above).

## 4. Conclusion

In the present paper it has been shown that the simple $[Cu_2(OH)_2F_4]^{2-}$ model complexes qualitatively reveal the same magneto-structural correlations that have been found for real doubly hydroxo-bridged Cu(II) dimers with bulky terminal ligands. Therefore, these model complexes are highly useful for describing these correlations in detail far beyond empirical relations derived from experimental data. For all applied structural changes of the complexes (i.e. bridging angle, out-of-plane angle of hydrogen, bending angle and Cu-O bonding distance) the transfer integral $H_{AB}$ in the antiferromagnetic contribution to the superexchange coupling constant has turned out to be the crucial quantity determining the magnetic behavior. The applied analytical approach supplies a simple scheme, based on orbital interactions, for calculating this transfer integral. The analytical results were in very satisfying agreement with fully numerical calculations of $H_{AB}$ and the magnetic coupling constant $J$. Moreover, with this approach the contributions to $H_{AB}$ of the single orbital interactions between Cu(3$d$) and the ligand orbitals have been explicitly calculated and discussed. This analysis supplies a much more detailed description of the magneto-structural correlations as those previously provided for the doubly hydroxo-bridged Cu(II)-dimers with fully numerical methods. The results do not only describe the magneto-structural correlations of the hydroxo-bridged complexes but may also serve as a basis for understanding the magnetic behavior of other compounds with similar structural variations.

**Acknowledgements**

This work has been financially supported by the Austrian Fonds zur Förderung der wissenschaftlichen Forschung (Project-No. P20503).

**Tables**

Table 1. The single contributions to $H_{AB}$, in cm$^{-1}$, for different bridging angles. *c*(*dd*) and *c*(*id*) give the contribution from direct Cu(3*d*)-Cu(3*d*) and metal-ligand interactions, respectively, where *i* is a ligand orbital.

| $\theta$ [°] | *c*(*dd*) | *c*(*zd*) | *c*(*sd*) | *c*(*Hd*) | *c*(*xd*) |
|---|---|---|---|---|---|
| 80  | 1189 | 4556 | -1165 | 60 | -5830 |
| 90  | 506  | 5971 | -1218 | 40 | -4942 |
| 95  | 328  | 6101 | -1226 | 36 | -4373 |
| 100 | 221  | 6678 | -1218 | 37 | -3745 |
| 110 | 102  | 7286 | -1169 | 50 | -2460 |

Table 2. The analytically calculated contributions, eqs.(6)-(9), to $H_{AB}$ (in cm$^{-1}$) for different out-of-plane angles of hydrogen. $c(dd)$ and $c(id)$ give the contribution from direct Cu(3$d$)-Cu(3$d$) and metal-ligand interactions, respectively, where $i$ is a ligand orbital.

| $\tau$ [°] | $c(dd)$ | $c(zd)$ | $c(sd)$ | $c(Hd)$ | $c(y'd)$ | $c(x'd)$ |
|---|---|---|---|---|---|---|
| 0  | 328 | 6101 | -1226 | 36  | 0     | -4373 |
| 20 | 328 | 6101 | -1239 | 37  | -435  | -4138 |
| 40 | 328 | 6101 | -1277 | 50  | -1511 | -3462 |
| 60 | 328 | 6101 | -1406 | 87  | -2901 | -2276 |
| 70 | 328 | 6101 | -1410 | 100 | -3439 | -1794 |

Table 3. The single contributions to $H_{AB}$, in cm$^{-1}$, for different bending angles $\gamma$ calculated with eqs.(6)-(9). $c(dd)$ and $c(id)$ give the contribution from direct Cu(3$d$)-Cu(3$d$) and metal-ligand interactions, respectively, where $i$ is a ligand orbital.

| $\gamma$ [°] | $c(dd)$ | $c(zd)$ | $c(sd)$ | $c(Hd)$ | $c(yd)$ | $c(xd)$ |
|---|---|---|---|---|---|---|
| 0 | 328 | 6101 | -1226 | 36 | 0 | -4373 |
| 3 | 339 | 6081 | -1225 | 36 | -20 | -4369 |
| 5 | 338 | 6064 | -1228 | 36 | -53 | -4372 |
| 10 | 377 | 5939 | -1227 | 36 | -211 | -4357 |
| 15 | 442 | 5721 | -1230 | 35 | -463 | -4339 |
| 20 | 547 | 5452 | -1233 | 35 | -793 | -4327 |
| 25 | 721 | 5139 | -1239 | 35 | -1194 | -4325 |
| 30 | 980 | 4757 | -1245 | 34 | -1647 | -4316 |

Table 4. The single contributions to $H_{AB}$, in cm$^{-1}$, as a function of the Cu-O and the Cu-Cu distance.

| $d$(Cu-Cu) [Å] | $d$(Cu-O) [Å] | $c(dd)$ | $c(zd)$ | $c(sd)$ | $c(Hd)$ | $c(xd)$ | $c(zd)+c(xd)$ |
|---|---|---|---|---|---|---|---|
| 2.51 | 1.70 | 1441 | 15438 | -5281 | 178 | -13527 | 1911 |
| 2.58 | 1.75 | 1074 | 12968 | -3952 | 127 | -10870 | 2098 |
| 2.65 | 1.80 | 796 | 10897 | -2957 | 92 | -8703 | 2194 |
| 2.73 | 1.85 | 594 | 9183 | -2212 | 67 | -6957 | 2226 |
| 2.88 | 1.95 | 328 | 6542 | -1226 | 36 | -4366 | 2176 |
| 3.10 | 2.10 | 130 | 4333 | -487 | 15 | -2164 | 2169 |
| 3.24 | 2.20 | 73 | 3616 | -261 | 10 | -1345 | 2271 |

**Figures**

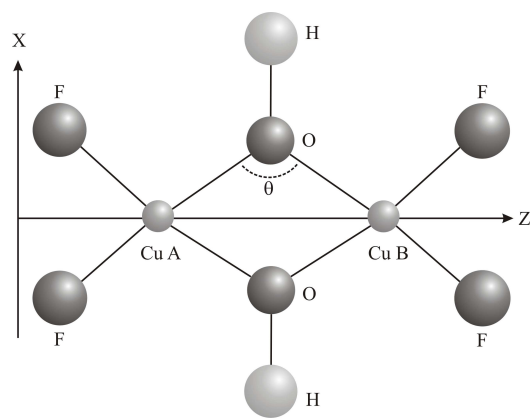

Figure 1.

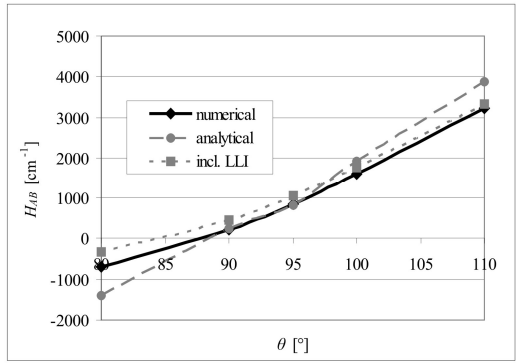

Figure 2.

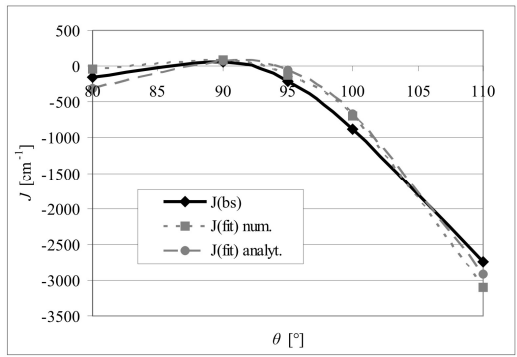

Figure 3.

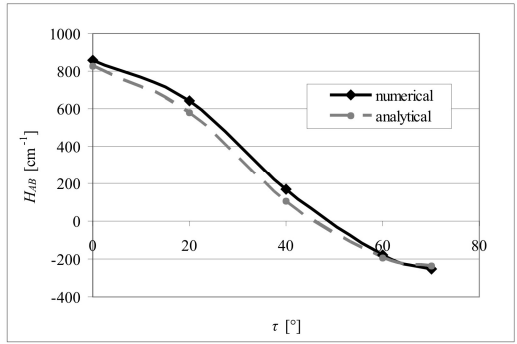

Figure 4.

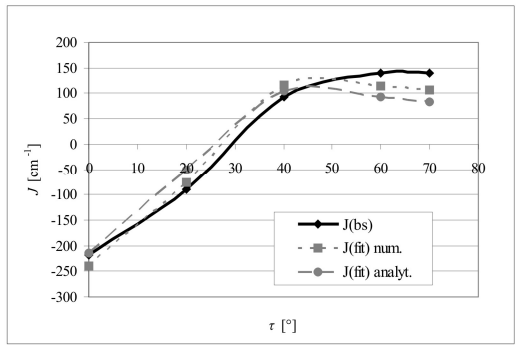

Figure 5.

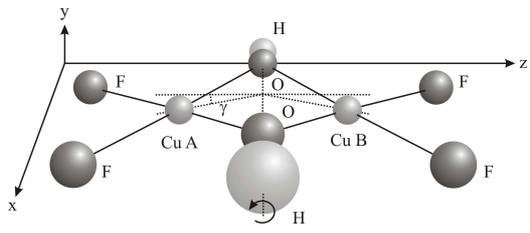

Figure 6.

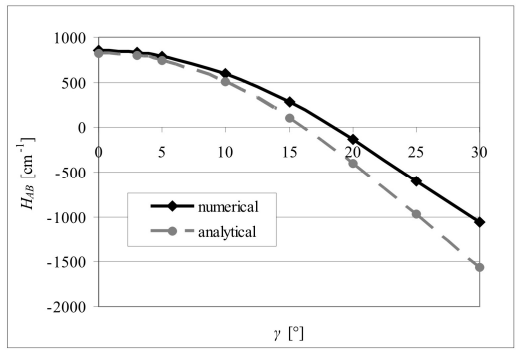

Figure 7.

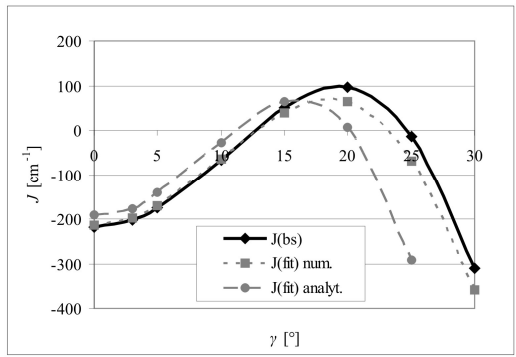

Figure 8.

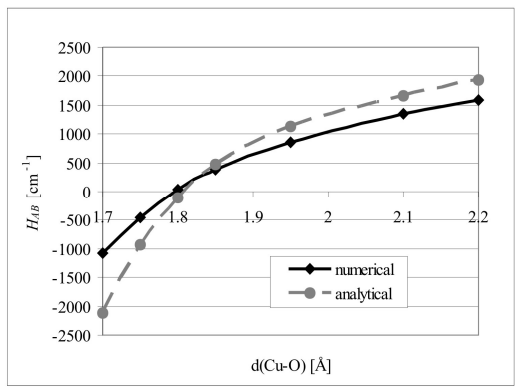

Figure 9.

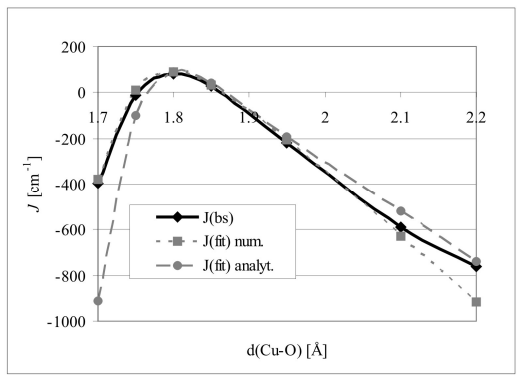

Figure 10.

**Figure captions**

Figure 1. Structure of the planar doubly-bridged $[Cu_2(OH)_2F_4]^{2-}$ complex.

Figure 2. $H_{AB}$, in cm$^{-1}$, as a function of the bridging angle $\theta$ calculated numerically with SCC-X$\alpha$, the analytical method and the analytical method supplemented with the ligand-ligand interactions (incl. LLI).

Figure 3. Comparison between the numerical coupling constants from broken symmetry calculations, $J(bs)$, and the results obtained with eq.(5) via inserting the numerical (num) $H_{AB}$ ($C = 100, f = 6500$) and the analytical (analyt) $H_{AB}$ ($C = 100, f = 10000$).

Figure 4. Comparison of the analytical and the SCC-X$\alpha$ (numerical) transfer integrals $H_{AB}$, in cm$^{-1}$, as a function of the out-of-plane angle of hydrogen $\tau$.

Figure 5. Comparison between the numerical coupling constants $J(bs)$ as a function of the out-of plane angle of hydrogen and the results obtained with eq.(5) via inserting the numerical (num) $H_{AB}$ ($C = 130, f = 4000$) and the analytical (analyt) $H_{AB}$ ($C = 110, f = 4200$).

Figure 6. Bent, roof shaped $[Cu_2(OH)_2F_4]^{2-}$ complex where the monomers are rotated into opposite directions by an angle $\gamma$ about the axis connecting the OH-bridges.

Figure 7. $H_{AB}$, in cm$^{-1}$, as a function of the bending angle $\gamma$ calculated numerically with SCC-X$\alpha$ and with the analytical method.

Figure 8. Comparison between the numerical coupling constants from the broken symmetry calculations, $J(bs)$, as a function of the bending angle $\gamma$ and the results obtained with eq.(5) via inserting the numerical (num) $H_{AB}$ ($C = 70, f = 5000$) and the analytical (analyt) $H_{AB}$ ($C = 70, f = 5000$).

Figure 9. $H_{AB}$, in cm$^{-1}$, as a function of the Cu-O distance calculated numerically with SCC-Xα and with the analytical method.

Figure 10. Comparison between the numerical coupling constants $J(bs)$ as a function of the Cu-O distance and the results obtained with eq.(5) via inserting the numerical (num) $H_{AB}$ ($C = 90, f = 5000$) and the analytical (analyt) $H_{AB}$ ($C = 90, f = 9000$).

**Graphical abstract**

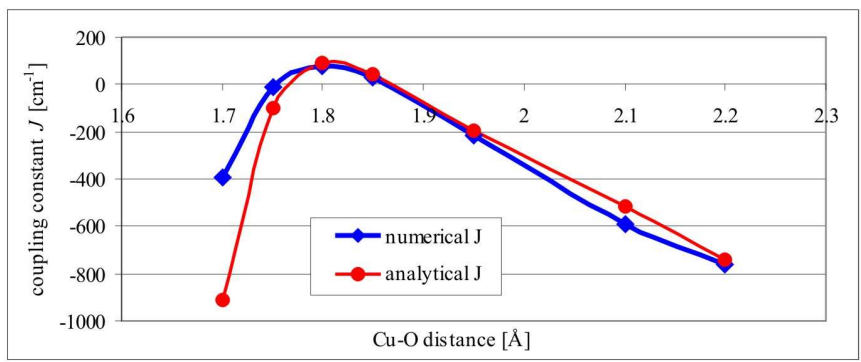